\documentclass[onecolumn,journal,12pt,draftclsnofoot]{IEEEtran}
\usepackage{amsmath}

\usepackage{units}
\usepackage{graphicx}
\usepackage{amssymb}

\usepackage[short]{optidef}

\usepackage{afterpage}
\usepackage{subfigure}
\usepackage{verbatim}
\usepackage{psfrag}
\usepackage{color}
\usepackage[table]{xcolor}
\usepackage{cite}
\usepackage{times}
\usepackage{wrapfig}
\usepackage{pgfplots}
\usepackage{multirow}
\usepackage{tikz}
\usepackage{setspace}
\usepackage{epsfig}
\usepackage{epstopdf}
\usepackage{footmisc}
\usepackage{fmtcount}
\usepackage{stackrel}
\usepackage{float}
\usepackage{enumerate}
\usepackage{caption}
\usepackage{hyperref}
\usepackage{hypcap}
\usepackage[nottoc]{tocbibind}
\usepackage[normalem]{ulem}
\usepackage{authblk}
\usepackage{tabularx}
\usepackage{array}
\usepackage{listings}

\usepackage{algorithm}
\usepackage{algpseudocode}

\usepackage{stackengine}
\setstackEOL{\\}
\usepackage{tabularx}
\makeatletter

\usepackage{babel}

\newcommand{\multiline}[1]{%
  \begin{tabularx}{\dimexpr\linewidth-\ALG@thistlm}[t]{@{}X@{}}
    #1
  \end{tabularx}
}
\makeatother

\usepackage{capt-of}

\begin{document}
\bstctlcite{IEEEexample:BSTcontrol}

\title{Deep Reinforcement Learning for RIS-Assisted FD Systems: Single or Distributed RIS?}


\author{Alice~Faisal,~\IEEEmembership{Graduate Member,~IEEE,}
        Ibrahim~Al-Nahhal,~\IEEEmembership{Senior Member,~IEEE,}
        Octavia~A.~Dobre,~\IEEEmembership{Fellow,~IEEE}
        and~Telex~M. N.~Ngatched,~\IEEEmembership{Senior Member,~IEEE}
\thanks{
The authors are with the Faculty of Engineering and Applied Science, Memorial University of Newfoundland, St. John’s, NL,
Canada, (e-mail: afaisal@mun.ca; ioalnahhal@mun.ca; odobre@mun.ca; tngatched@grenfell.mun.ca).
} }

\maketitle
\begin{abstract}
This paper investigates reconfigurable intelligent surface (RIS)-assisted full-duplex multiple-input single-output wireless system, where the beamforming  {and} RIS phase shifts are optimized to maximize the sum-rate for both single and distributed RIS  deployment schemes. The preference of using the single or distributed RIS deployment scheme is investigated through three practical scenarios based on the links’ quality. 
 {The closed-form}  {solution is} derived to optimize the beamforming vectors and a novel deep reinforcement learning (DRL) algorithm is proposed to optimize the RIS phase shifts. Simulation results illustrate that the choice of the deployment scheme depends on the scenario and the links' quality. It is further shown that the proposed algorithm significantly improves the sum-rate compared to the non-optimized scenario in both single and distributed RIS  deployment schemes. Besides, the proposed beamforming derivation achieves a remarkable improvement compared to the approximated derivation in previous works.  Finally, the complexity analysis confirms that the proposed DRL algorithm reduces the computation complexity compared to the DRL algorithm in the literature.

\end{abstract}

\IEEEpeerreviewmaketitle
\section{Introduction}

\IEEEPARstart{R}{ecently}, {the} reconfigurable intelligent surface (RIS) technology has been proposed as a key enabler to meet the demands of future technologies \cite{9350614, Ibra2}.
{RIS} is a meta-surface  {consisting} of low-cost passive  elements that can be programmed to turn the random nature of wireless channels into a partially deterministic space  {to improve} the propagation of wireless  {signals} \cite{RIS_survay}. In addition to the RIS technology, full-duplex (FD) transmission has been regarded as a potential approach to increase  {the spectral efficiency} of wireless systems  {by enabling simultaneous transmission and reception} \cite{full-duplex,full-duplex2}.




{Incorporating RIS into FD communications} can {provide} new degrees of freedom, facilitating ultra spectrum-efficient communication systems \cite{RIS_Survay2}. A number of existing works have studied  {RIS-assisted FD wireless networks} \cite{shen2020beamforming,exact_FD,Dist_Opt_FD}. The  {works} in \cite{shen2020beamforming,exact_FD} considered alternating optimization (AO) techniques to optimize the RIS  {phase shifts} in FD systems. The authors in \cite{Dist_Opt_FD} considered {a multi RIS-assisted} FD system to maximize the weighted system sum-rate, where the non-convex problem was addressed using the AO approach.



 {The} above works that used AO techniques exhibit both loss of optimality and high computational complexity. Deep reinforcement learning (DRL) has emerged as a powerful approach to optimize the RIS phase shifts by overcoming the practical implementation problems of  {AO techniques.} {Furthermore, DRL approaches enable addressing mathematically intractable nonlinear problems directly, without the need of prior relaxations requirements.}
The work in \cite{A_letter} {proposed} a DRL algorithm to maximize the rate, where both half-duplex and FD operating modes are considered together.  {However, only a single RIS deployment was considered.}
 {The rapid changes in dynamic environments can obliterate/annihilate the RIS deployment benefits when the corresponding link is blocked/weak.}
 In such cases, deploying distributed power-efficient RISs can cooperatively enhance the coverage of the system by providing multiple paths of received signals. {Moreover, the computational complexity can be further reduced.}
 {To the best of the authors’ knowledge, utilizing DRL for investigating the performance of single and distributed RIS deployment schemes {in FD multiple-input} single-output
(MISO) systems has not yet been considered in the literature.   {Our} contributions are summarized as follows:}


\begin{itemize}
    \item Three practical scenarios are considered to investigate the sum-rate performance of deploying {a} single or distributed RIS  in {a FD-MISO system.}

    \item {A closed-form solution is derived to optimize the transmit beamformers, which provides a remarkable improvement in the sum-rate compared to the state-of-the-art approximated derivation in \cite{A_letter}}.
    \item {An improved DRL algorithm is proposed to optimize the RIS phase shifts for both deployment schemes, which achieves a significant improvement in the sum-rate compared to the non-optimized scenarios.}
    \item {The proposed DRL algorithm {provides} a considerable reduction in the computational complexity compared to the DRL algorithm in \cite{A_letter}.}
    \item The complexity analysis and Monte Carlo simulations support the findings.
\end{itemize}

The rest of {this} letter is organized as follows: Section \ref{system_model} presents the system model and problem formulation. The proposed DRL algorithm is introduced in Section \ref{solution}.  {Simulation} results and {conclusions} are presented in Sections \ref{simulations} and \ref{conclusion}, respectively.



\section{System Model and Problem Formulation} \label{system_model}

\begin{figure}[t!]
    \centering
    \subfigure[Scenario 1]{\includegraphics[width=0.4\textwidth]{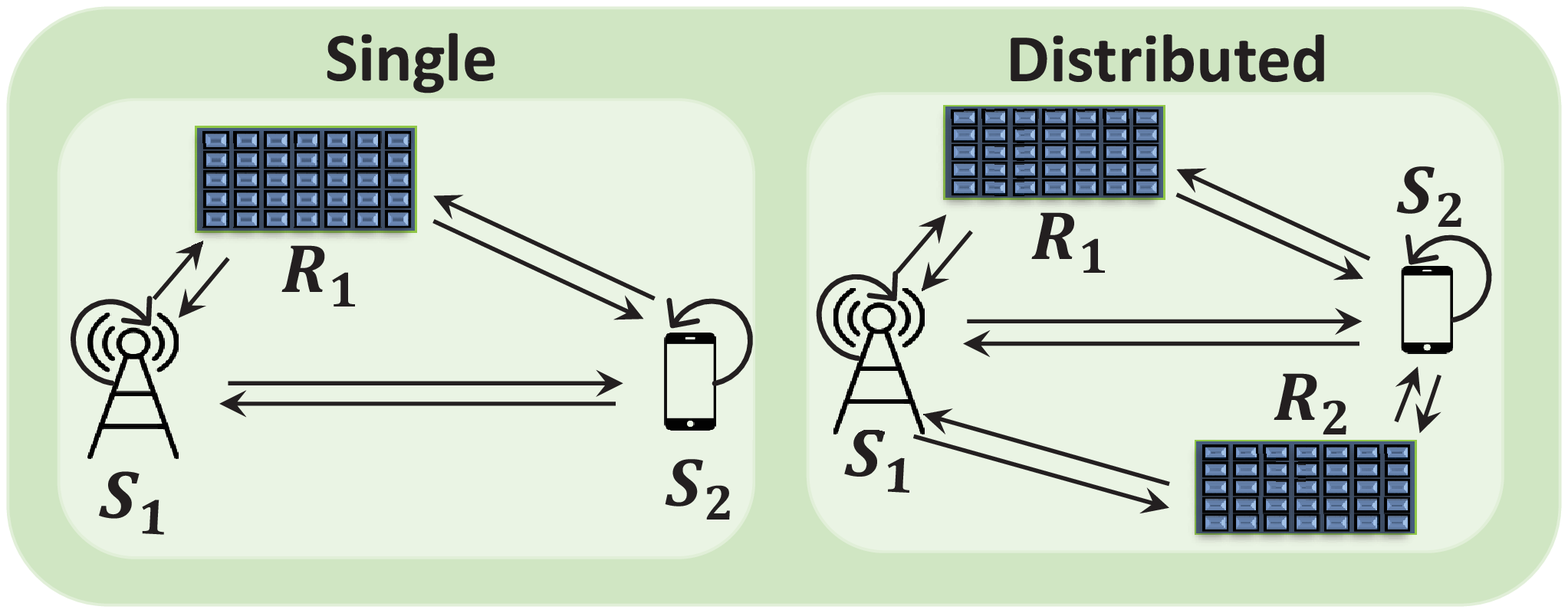}} 
    \subfigure[Scenario 2]{\includegraphics[width=0.4\textwidth]{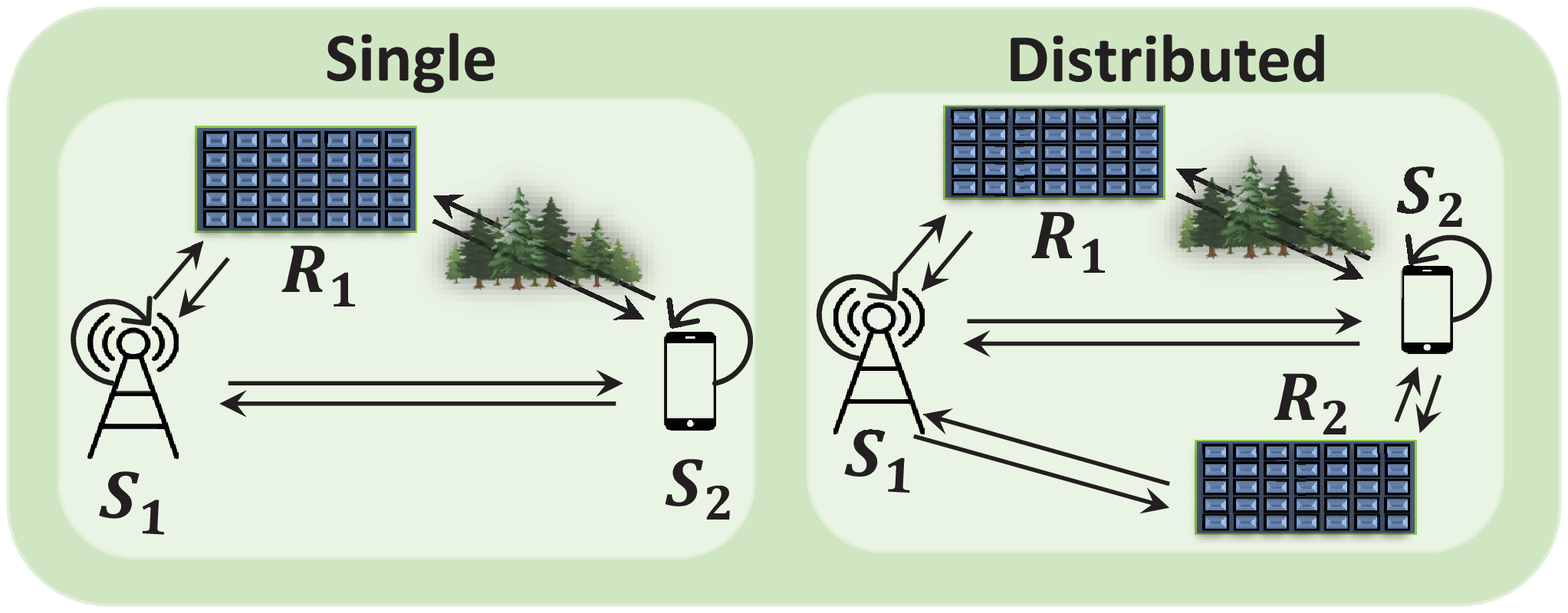}}
    \subfigure[Scenario 3]{\includegraphics[width=0.4\textwidth]{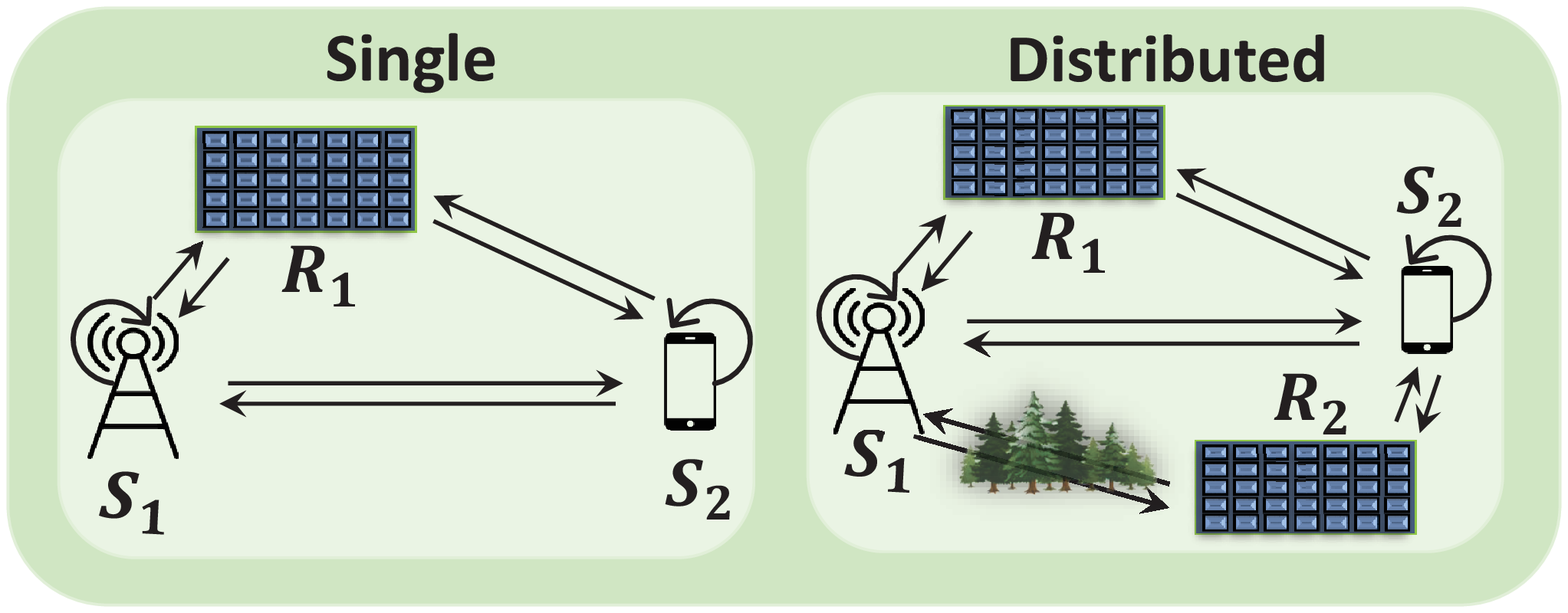}}
    \caption{RIS-assisted FD MISO system. }
    \label{fig:foobar}
\end{figure}

Consider an RIS-assisted FD MISO system, where single and distributed RIS deployment schemes are investigated. $S_1$ and $S_2$ represent the base station (BS) and user equipment (UE), respectively. {Both the BS and UE are equipped with $M$ transmit antennas and one receive antenna.} {The $r$-th RIS, $R_r$, consists of}
$N_r$ programmable reflecting elements.
Note that the total number of elements for both deployment schemes is defined as {$N = N_r  \Lambda$} to ensure the same number of RIS elements for all scenarios, {where $  \Lambda$ is the number of RISs.} As illustrated in Fig. \ref{fig:foobar}, three scenarios are investigated based on the links' quality.  {In the first scenario,} the single and distributed RIS  {deployment schemes} have strong line-of-sight (LoS) components in all links.  Scenarios 2 and 3   {assume} that the links of  {$R_1$}-$S_2$ and $S_1$-$R_2$ are weak due to obstacles, respectively. 
{It is worth noting that from a practical point of view, it is more probable that the longer distance links (i.e., $R_1$-$S_2$ and $S_1$-$R_2$) may experience blockage since the short-distance links are planned deployment links. It also ensures a fair comparison between the two deployment schemes as the RIS benefits are embraced in all scenarios.}

Given $\bar{i} = 3 - i$   {$\forall \hspace{0.2em}i \in \{1,2\}$}, let $\mathbf{H}_{{S_{\bar{i}}R_r}}\in\mathbb{C}^{N_r\times M}$, $\mathbf{h}^H_{R_rS_i}\in\mathbb{C}^{1\times N_r}$, and $\mathbf{h}^H_{{S_{\bar{i}} S_{i}}}\in\mathbb{C}^{1\times M}$ denote the channel coefficients of the $S_{\bar{i}}$-$R_r$,  {$R_r$}-$S_i$, and $S_{\bar{i}}$-$S_i$ links, respectively. The self-interference (SI) channels of both the BS and UE are denoted by $\mathbf{h}^H_{S_iS_i} \in\mathbb{C}^{1\times M}$. Hence, the noisy received signal,  {$y_i$, is}

\begin{multline}
    y_i =\Bigl(\sum_{r \in  \Lambda} \mathbf{h}_{R_r S_i}^H\boldsymbol{\Theta}_r\mathbf{H}_{S_{\bar{i}}R_r} +  \mathbf{h}_{{S_{\bar{i}} S_{i}}}^H\Bigr)\mathbf{w}_{\bar{i}}x_{\bar{i}}  +   \mathbf{h}_{{S_{i} S_{i}}}^H \mathbf{w}_ix_i +   n,  \\ i = 1, 2, \hspace{2mm}   \Lambda = \begin{cases} 1 & \text{Single RIS}  \\  
    2 & \text{Distributed RIS},
    \end{cases}
    \label{FD_received}
\end{multline}


\noindent where $n\sim\mathcal{CN}(0,\sigma^2)$   {denotes} the additive white {complex} Gaussian noise {with zero-mean and {variance $\sigma^2$}.} The diagonal matrix $\boldsymbol{\Theta}_r = \text{diag}\bigl(e^{j\varphi_{r1}}, \cdots,e^{j\varphi_{rn}},\cdots,e^{j\varphi_{rN_r}}\bigr)\in\mathbb{C}^{N_r\times N_r}$  {represents} the phase shifts of  {$R_r$}, where  $\varphi_{rn}\in[-\pi,\pi)$ is the phase shift introduced by the $n$-th reflecting element. The source node, $S_i$, employs an active beamforming $\mathbf{w}_i\in\mathbb{C}^{M\times 1}$ to transmit the information signal, $x_i$, with $\mathbb{E}\{|x_i|^2\}=1$, where $\mathbb{E}\{\cdot\}$ denotes the expectation operation. 

Based on (\ref{FD_received}), the received signal-to-interference plus-noise ratio,  {$\gamma_i$, and achievable rate, $\mathcal{R}_i$, measured in bit per second per Hertz (bps/Hz), are respectively given as}



  \begin{multline} 
    \gamma_i =  \frac{\left|\left( \sum_{r \in  \Lambda} \mathbf{h}^H_{R_rS_{{i}}} \boldsymbol{\Theta}_r\mathbf{H}_{S_{\bar{i}}R_r} + \mathbf{h}^H_{S_{\bar{i}}S_i}\right)\mathbf{w}_{\bar{i}}\right|^2}{|\mathbf{h}^H_{S_iS_i} \mathbf{w}_i|^2 + \sigma^2}, \\ i = 1, 2, \hspace{2mm}  \Lambda = \begin{cases}  1 & \text{Single RIS}  \\  
     2 & \text{Distributed RIS},
    \end{cases}
    \label{FD:gamma}
\end{multline} 

\noindent and

\begin{equation}
       \mathcal{R}_i = \text{log}_2\left(1+ \gamma_i \right).
       \label{FD:reward}
\end{equation}




\noindent  {The} objective is to maximize the sum-rate by optimizing the beamformers and RIS phase shifts,  {and} is formulated as


\begin{maxi!}
{\mathbf{w}_{{i}}, \bar{\boldsymbol{\Theta}}}{\sum_{i=1}^2 \mathcal{R}_i} 
{}{\text{(P1)}\quad} \label{eq:obj}
\addConstraint{{ -\pi \leq \varphi_{rn} \leq \pi, \hspace{0.2em} n = 1, \cdots , N_r}} \label{eq:IntRIS1}
\addConstraint{{ ||\mathbf{w}_{{i}}||^2 \leq P_{\text{max}}, \hspace{0.2em} i = 1, 2. }} \label{eq:IntRIS2}
\end{maxi!}

\noindent  {Here,} $\bar{\boldsymbol{\Theta}} = \text{diag}\bigl(\boldsymbol{\Theta}_1, \boldsymbol{\Theta}_2\bigr)$ is a block matrix  {whose} diagonal entries contain the phase shifts of the two RISs  {for the} distributed RIS,  {and} $\bar{\boldsymbol{\Theta}} = \text{diag}\bigl(\boldsymbol{\Theta}_1)$  {when a single RIS is considered.} $P_{\text{max}}$ is the maximum transmitted power of $S_{{i}}$.
{It is worth noting that (P1) is challenging to  {solve due to the non-convexity of the objective function and constraints. Thus,} an efficient solution is proposed which decouples the problem into two sub-problems.}

\section{Proposed Solution}
\label{solution}
{This section proposes a novel algorithm to solve (P1).  {First}, a closed-form solution is derived to optimize the transmit beamformers, $\mathbf{w}^\ast_{{i}}$, for a fixed $\boldsymbol{\bar{\Theta}}$. Then, the RIS phase shifts, $\boldsymbol{\bar{\Theta}}$, are {obtained} using the proposed DRL algorithm. This process is repeated until {$\boldsymbol{\bar{\Theta}}^\ast$ and $\mathbf{w}^\ast_{{i}}$} converge. In what follows, more details about the two-step solution are provided.}


\subsection{Beamformers Optimization for a Given $\boldsymbol{\bar{\Theta}}$}


The mutual information $I(s;y)$ with an arbitrary input probability distribution $p(s)$ for a channel with input $s$, output $y$, and a transition probability of $p(y|s)$ is given by

\begin{equation}
    I(s;y) = \max_{q(s|y)} \mathbb{E} \bigl[ \log\left(q(s|y)\right) - \log\left(p(s)\right) \bigr], 
    \label{mutualInf}
\end{equation}

\noindent  {where  {the optimal $q^\ast(s|y)$ is the posterior probability \cite{exact_FD}, and is expressed as $q^\ast(s|y) = \frac{p(s)p(y|s)}{p(y)} \triangleq p(s|y)$}.}
Based on \eqref{mutualInf}, the  {achievable rate} of $S_i$ is

\begin{equation}
\mathcal{R}_i = \max_{q(s_i|y_{\bar{i}})} \mathbb{E} \bigl[ \log\left(q(s_i|y_{\bar{i}})\right) - \log\left(p(s_i)\right) \bigr],
\end{equation}

\noindent where the input probability distribution $p(s_i)$ is $\mathcal{CN}(0,1)$ and the channel transition probability $p(y_{\bar{i}}| s_i)$ is obtained from \eqref{FD_received}. 
According to \cite{estbook}, $p(s_i|y_{\bar{i}})$ follows the complex Gaussian distribution of $\mathcal{CN}(f^\ast_{\bar{i}} y_{\bar{i}},\Sigma^\ast_{\bar{i}})$. $\Sigma^\ast_{\bar{i}}$ is defined as $\Sigma^\ast_{\bar{i}} = 1 - f^\ast_{\bar{i}}  b_{\bar{i}}$, where $f^\ast_{\bar{i}}$ and $b_{\bar{i}}$ are respectively expressed as

\begin{equation}
 f^\ast_{\bar{i}} =  \frac{b_{\bar{i}}}{b^2_{\bar{i}} + |\mathbf{h}_{{S_{\bar{i}} S_{\bar{i}}}}^H \mathbf{w}_{\bar{i}}|^2},
 \end{equation}
 \begin{equation}
 b_{\bar{i}} = \left|\left( \sum_{r \in  \Lambda} \mathbf{h}^H_{R_rS_{\bar{i}}} \boldsymbol{\Theta}_r\mathbf{H}_{S_{{i}}R_r} + \mathbf{h}^H_{S_{{i}}S_{\bar{i}}}\right)\mathbf{w}_{{i}}\right|^2. 
 \end{equation}
 
  To this end, \eqref{eq:obj} in (P1) can be re-expressed as
  
\begin{maxi}
{\mathbf{w}_{{i}}, \bar{\boldsymbol{\Theta}}, f_{{i}}, \Sigma_{{i}}}{\sum_{i=1}^2 \mathbb{E} \bigl[ \log(p(s_i|y_{\bar{i}})) - \log(p(s_i))\bigr]}
{}{}.
\label{Obj2}
\end{maxi}


   \noindent Let $\boldsymbol{\alpha}_{\bar{i}} =\sum_{r \in  \Lambda} \mathbf{h}^H_{R_rS_{\bar{i}}} \boldsymbol{\Theta}_r\mathbf{H}_{S_{{i}}R_r} + \mathbf{h}^H_{S_{{i}}S_{\bar{i}}}$ and $b_{\bar{i}} = |\boldsymbol{\alpha}_{\bar{i}} \mathbf{w}_i|^2$. The expectation term in \eqref{Obj2} is calculated as
   \begin{multline}
       \mathbb{E} \bigl[ \log(\mathcal{CN}(f_{\bar{i}} y_{\bar{i}},\Sigma_{\bar{i}})) - \log({\mathcal{CN}(0,1)})\bigr] \\ =  \exp{\left(f_{{i}}y_{\bar{i}} + \frac{\Sigma_{\bar{i}}}{2}\right)} - \exp{\left(\frac{1}{2}\right)} \\
       = - \frac{1}{2} f_{\bar{i}} |\boldsymbol{\alpha}_{\bar{i}} \mathbf{w}_i|^2 + \mathbf{w}_i\left(f_{\bar{i}}  \boldsymbol{\alpha}_{\bar{i}} + f_{{i}} \mathbf{h}^H_{S_{{i}}S_i}\right).
       \label{BFExp}
    \end{multline}
    
     
     \noindent  {Furthermore, let $\boldsymbol{\beta}_{\bar{i}} \hspace{-1mm} = \hspace{-1mm} f_{\bar{i}}  \boldsymbol{\alpha}_{\bar{i}} + f_{{i}} \mathbf{h}^H_{S_{{i}}S_i}$. Thus, \eqref{BFExp} can be defined as a convex quadratically constrained quadratic program:}

\begin{equation}
    - \frac{1}{2} f_{\bar{i}} |\boldsymbol{\alpha}_{\bar{i}} \mathbf{w}_i|^2 + \mathbf{w}_i \boldsymbol{\beta}_{\bar{i}},
\end{equation}

\noindent where its solution can be derived as

\begin{equation}
    \mathbf{w}^\ast_{{i}} =  (v^\ast + f_{\bar{i}} \boldsymbol{\alpha}_{\bar{i}} \boldsymbol{\alpha}^H_{\bar{i}})^{-1} \boldsymbol{\beta}_{\bar{i}}.
    \label{beamformingFD}
\end{equation}

\noindent  Here, $v^\ast$ is the optimal dual {Lagrangian} variable associated with the power constraint  {and} is found by performing a bisection search over the interval $\left[0, \sqrt{
\boldsymbol{\beta}_{\bar{i}}^T \boldsymbol{\beta}_{\bar{i}}}/\sqrt{P_{\text{max}}}\right]$ \cite{boyd2004convex}.



\subsection{Phase Shift Optimization for a {Given} $\mathbf{w}_{{i}}$ and $\mathbf{w}_{\bar{i}}$}


Model-free RL can be employed to address a decision-making problem by learning the optimal solution in dynamic environments. Therefore, the RIS-assisted FD MISO system represents the DRL environment and  {the RIS} controller represents the DRL agent. At each time step $t$, the agent observes the current state,  {$s_t$}, from the environment, takes an action, {$a_t$}, based on a policy, $\tilde{\pi}$, receives a reward, {$r_t$}, of executing $a_t$, and transitions to a new state $s_{t+1}$. The key elements of DRL are defined as follows: The \textit{state space} at time step $t$, includes $\varphi_{rn} \hspace{0.1em}  {\forall \hspace{0.2em} n} = 1, \cdots, N_r$ and the corresponding $\sum_{i=1}^2$ $\mathcal{R}_i$  at time step $t-1$, i.e., $s_t \hspace{-1mm} = \hspace{-1mm} \left[\sum_{i=1}^2 \mathcal{R}^{(t-1)}_i, \varphi^{(t-1)}_{r1}, \cdots,\varphi^{(t-1)}_{rn},\cdots,\varphi^{(t-1)}_{rN_r}\right]$. The \textit{action space} at time step $t$ is expressed as $a_t = \left[\varphi^{(t)}_{r1}, \cdots,\varphi^{(t)}_{rn},\cdots,\varphi^{(t)}_{rN_r}\right]$, and the \textit{reward} at time step $t$ is $r_t = \sum_{i=1}^2 \mathcal{R}^{(t)}_i$.

The goal of a RL agent is to learn a policy that maximizes the expected cumulative discounted reward from the start state, as: $J(\tilde{\pi}) = \mathbb{E}\left[R_1 | \tilde{\pi}\right]$. The policy gradient based algorithms can be used to learn the optimal policy for continuous $a_t$. In particular, the proposed algorithm aims at maximizing the return by training deep neural networks (DNN) to approximate the Q-value function. It is based on the \textit{actor-critic} approach, which consists of two DNN models: \textit{actor}, $\mu(s_t|\boldsymbol{\theta}_\mu)$, and \textit{critic}, $Q(s_t,a_t|\boldsymbol{\theta}_q)$, where $\boldsymbol{\theta}$ represents the DNN parameters. The actor takes the state as an input and outputs  $a_t = \mu(s_t|\boldsymbol{\theta}_\mu) + \xi$, where $\xi$ is a random process that is added to the actions for exploration, representing the policy network. The critic takes $s_t$ and $a_t$ as an input and outputs the Q-value, representing {the evaluation} network \cite{lillicrap2019continuous}.

At the initialization stage, four networks are generated, i.e., target and evaluation DNN. The target networks are generated by making a copy of the actor and critic evaluation NNs, $\mu'(s_t|\boldsymbol{\theta}_{\mu'})$ and $Q'(s_t,a_t|\boldsymbol{\theta}_{q'})$. The experience replay with memory $D$ is built to reduce the correlation of the training samples. During each episode, all the channel state information {is} obtained. Then, the agent takes $a_t$ generated by the actor network, calculates the $r_t$, and transitions to $s_{t+1}$. The experience is then stored $(s_t, a_t, r_t, s_{t+1})$ into $D$, and the critic evaluation network randomly samples a minibatch transitions, $N_B$, to calculate the target value $y_j$, as

\begin{equation}
    y_j = r_j + \rho Q'(s_{j+1},\mu'(s_{j+1}|\boldsymbol{\theta}_{\mu'})|\boldsymbol{\theta}_{q'}),
    \label{target eq}
\end{equation}

\noindent where $\rho \in (0,1]$ is the discount factor. The actor and critic NN parameters, $\boldsymbol{\theta}_\mu$ and $\boldsymbol{\theta}_q$, are updated using the stochastic gradient descent and policy gradient, respectively, as 

\begin{equation}
    L = \frac{1}{N_B} \sum_j \left(y_j - Q(s_j,a_j|\boldsymbol{\theta}_q)\right)^2,
    \label{SGD eq}
\end{equation}

\noindent and 
\begin{equation}
    \nabla_{\boldsymbol{\theta}_\mu} = \frac{1}{N_B}\sum_j \nabla_a Q(s,a|\boldsymbol{\theta}_q)|_{s=s_j,a = \mu(s_j)} \nabla_{\boldsymbol{\theta}_\mu} \mu(s|\boldsymbol{\theta}_\mu)|_{s_j}.
    \label{pg eq}
\end{equation}

\noindent Finally, the target NN parameters are updated using a soft update coefficient, $\tau$, as
\begin{equation}
    \boldsymbol{\theta}_{q'} \longleftarrow \tau \boldsymbol{\theta}_{q} + (1-\tau) \boldsymbol{\theta}_{q'},
    \label{taR1 eq}
\end{equation}
\begin{equation}
    \boldsymbol{\theta}_{\mu'} \longleftarrow \tau \boldsymbol{\theta}_{\mu'} + (1-\tau) \boldsymbol{\theta}_{\mu'}. 
    \label{TAR2 eq}
\end{equation}

\begin{figure}[t]
  \centering
  \includegraphics[scale=0.45]{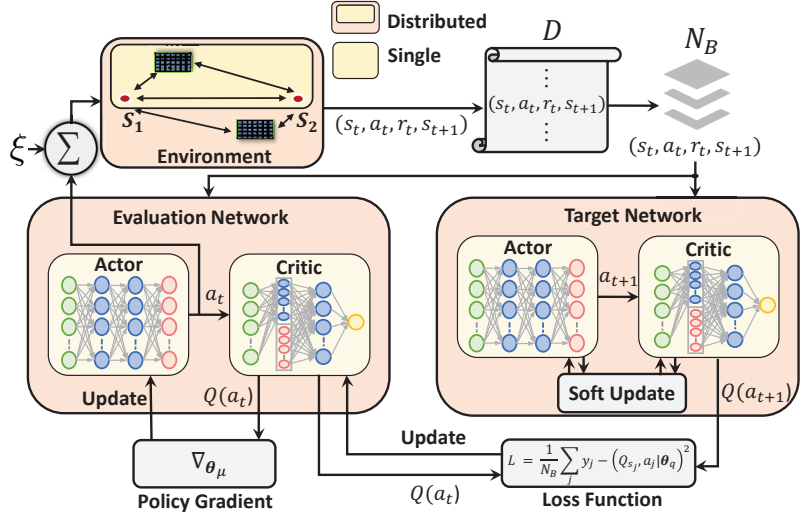}
  \caption{The proposed DRL algorithm structure. }
  \label{fig:DDPG}
\end{figure}

\noindent This process is repeated for $K$ and $T$ until {convergence is reached.} The structure of the proposed DRL algorithm is illustrated in Fig. \ref{fig:DDPG} and summarized in Algorithm 1.

\setlength{\textfloatsep}{2pt}
\begin{algorithm}[h!]
\caption{Proposed DRL {algorithm. }}
\begin{algorithmic}[1]
\Require $\boldsymbol{\theta}_\mu$ and $\boldsymbol{\theta}_q$ with random weights, $D$, $\rho$, $\tau$, learning rate $\nu$, $\boldsymbol{\theta}_{\mu'} \leftarrow \boldsymbol{\theta}_\mu$ and $\boldsymbol{\theta}_{q'} \leftarrow \boldsymbol{\theta}_q$;
\Repeat 
\State \multiline{Collect the channels of the $k$-th episode;}
\State \multiline{Randomly initialize $\varphi_{rn} \hspace{0.2em}  {\forall \hspace{0.2em}} n = 1, \cdots, N_r$;}
\State{Calculate {$\mathbf{w}_{\bar{i}}$} using \eqref{beamformingFD}};
\State Initialize  $\xi\sim\mathcal{CN}(0,0.1)$;
\Repeat 
\State \multiline{Obtain $a_t = \mu(s_t|\boldsymbol{\theta}_\mu) + \xi$ from the actor network and reshape it;}
\State{Repeat \textbf{Line} \#4;}
\State{Observe the new state, $s_{t+1}$, given $a_t$; }
\State \multiline{Store ($s_t,a_t,r_{t},s_{t+1}$) in $D$;}
\State  {\multiline{When $D$ is full, sample a minibatch of $N_B$ transitions  {($s_j,a_j,r_{j},s_{j+1}$) randomly} from $D$;}}
\State{\multiline{Compute the target value from \eqref{target eq};}}
\State \multiline{Update the critic using \eqref{SGD eq};}
\State \multiline{Update the actor using \eqref{pg eq};}
\State \multiline{Update the target NNs using \eqref{taR1 eq} and \eqref{TAR2 eq};}
 \Until{$t = T$;} 
 \Until{$k = K$;}
\Ensure{Optimal action that corresponds to the optimal  {$\boldsymbol{\bar{\Theta}}^\ast$}.}
\end{algorithmic}
\label{al:BCD}
\end{algorithm}

\subsection{Proposed DNN Design} \label{Complexity}

{The proposed DNN models are designed {as feedforward} fully connected NNs. The proposed algorithm contains four NNs (actor and critic for {each evaluation} and target network). Each NN has an input layer, two hidden layers and output layer, as {shown} in Fig.  \ref{fig:DDPG}. }
The input layer of the actor and critic networks contains $N + 1$ neurons (i.e., size of $s_t$). The input of the actor is passed to two hidden layers,  {each having} $\psi_i$ neurons, where $\psi_i$ is the number of neurons of the $i$-th layer. {On the other hand, the input of the critic network is passed to the first hidden layer that is concatenated {with $a_t$} (i.e., size of $\psi_i + N$), and then passed to the second hidden layer.} {The two hidden layers for each of the actor and critic networks use the \textit{ReLU} activation function whereas the output layer of the actor network uses the \textit{tanh} activation function.} The output layer of the actor and critic networks contains $N$ neurons (i.e., size of $a_t$) and one neuron (i.e., Q-value), respectively.


\subsection{Complexity Analysis}

The computational complexity of the proposed DRL algorithm is analyzed in terms of the number of NN parameters $C_\mathcal{P}$ required to be stored, real additions $C_\mathcal{A}$, and real multiplications $C_\mathcal{M}$. It is worth noting  that, for simplicity, each activation function is considered to cost one real addition. Henceforth, the complexity for the proposed DRL algorithm based on the NNs design is given as

\begin{align}
    & C_\mathcal{P} = 2\left(\sum_{i=1}^3 (\psi^{\text{A}}_{i}+1)\psi^{\text{A}}_{i+1} + \sum_{i=1}^3 (\psi^{\text{C}}_{i}+1)\psi^{\text{C}}_{i+1}\right), \\
    & C_\mathcal{M} = 2\left(\sum_{i=1}^3 \psi^{\text{A}}_i \psi^{\text{A}}_{i+1} + \sum_{i=1}^3 \psi^{\text{C}}_i \psi^{\text{C}}_{i+1}\right), \\
    & C_\mathcal{A} = 2\left(\sum_{i=1}^3 \psi^{\text{A}}_i \psi^{\text{A}}_{i+1} + \sum_{i=1}^3 \psi^{\text{A}}_{i+1} + \sum_{i=1}^3 \psi^{\text{C}}_i \psi^{\text{C}}_{i+1} + \sum_{i=1}^3 \psi^{\text{C}}_{i+1}\right),
\end{align}

\noindent where the actor and critic networks are expressed {through the superscripts} A and C, respectively. The complexity reduction of using the proposed DRL algorithm over the algorithm in \cite{A_letter}  for the single RIS-assisted FD system is

\begin{equation}
    \text{Reduction} =  1 - \frac{ \left\{C^{\text{A}}_\chi + C^{\text{C}}_\chi\right\}_{\text{Proposed}}}{\left\{C^{\text{A}}_\chi + C^{\text{C}}_\chi \right\}_{\text{\cite{A_letter}}}}, \hspace{0.2em} \chi \in \{\mathcal{P}, \mathcal{A}, \mathcal{M}\}.
\end{equation}







\section{Simulation Results}\label{simulations}

\begin{figure}[t]
    \centering
\includegraphics[scale=0.4]{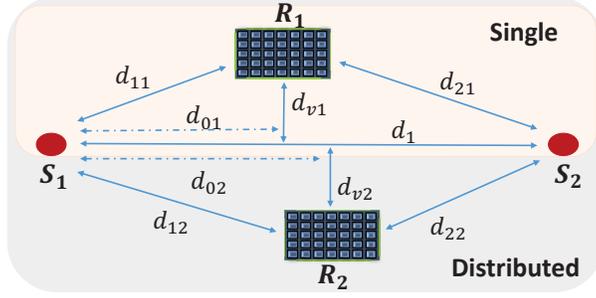}
    \caption{Simulation setup. }
    \label{fig:setup}
\end{figure}

Figure \ref{fig:setup} {illustrates} the simulation setup, where the considered parameters are: $d_{v1} = d_{v2} = \unit[2]{m}$ and $d_1 = \unit[50]{m}$. The distances between the links are: $d_{11} = \unit[\sqrt{d^2_{01}+d^2_{v1}}]{m}$,  $d_{12} = \sqrt{d^2_{02}+d^2_{v2}}$ m, $d_{21} = \sqrt{(d_1-d_{01})^2+d^2_{v1}}$ m, and $d_{22} = \sqrt{(d_1-d_{02})^2+d^2_{v2}}$ m.  The path loss (PL) at distance  {$d_{ir}$} is modeled as
$\text{PL}$ = $PL_0 - 10 \zeta \text{log}_{10}\left(\frac{d_{ir}}{ {D_r}}\right)$ \cite{DRLmain}, where $PL_0$ is the PL at a reference distance $ {D_r}$ and $\zeta$ is the PL exponent, in which $PL_0 =  {\unit[-35.6]{dB}}$ and $ {D_r} = \unit[1]{m}$.
The channels are modeled as {Rayleigh fading} whenever a blocking element exists. Otherwise, the channels are modeled as Rician with a factor of $10$. The PL exponents of the $S_1$-$S_2$, $S_1$-$R_r$, and $S_2$-$R_r$ channels are set to $\zeta_{\text{BU}} = 4$, $\zeta_{\text{BR}} = 2.1$,  and $\zeta_{\text{UR}} = 2.2$, respectively \cite{Dist_Opt_FD}. The PL of the SI channels is $\unit[-95]{dB}$. The total transmit power is $P = \unit[15]{dBm}$, while the noise power is $\sigma^2$ = $\unit[-80]{dBm}$ \cite{shen2020beamforming}. 


The parameters of the proposed DRL are as follows: $T = 800$, $K = 500$, $N_B = 16$, {$\nu_{\text{A}} = 0.0001$, $\nu_{\text{C}} = 0.0002$}, decaying rate = $0.0001$, $\rho = 0.99$, $\tau = 0.001$,  {and} $D = 50000$. Both actor and critic networks use {the Adam} optimizer for updating the parameters. {The number of neurons of the hidden layers are, $\psi_1 = 100$ and $\psi_2 = 45$.}
{To validate the performance of the proposed algorithm, it is compared with the non-optimized {scheme}, referred to as {random phase shifts.} Furthermore, it is compared with the algorithm in \cite{A_letter} for the single RIS-assisted FD system to show the superiority of the proposed beamforming derivations over the approximated derivations in \cite{A_letter}.} To ensure a fair comparison, it is assumed {that $N$} is the same for both deployment {schemes}. Hence, each RIS in the distributed scheme has half the number of elements of the single {scheme}. 

\begin{figure}[t]
    \centering
    \includegraphics[scale=0.5]{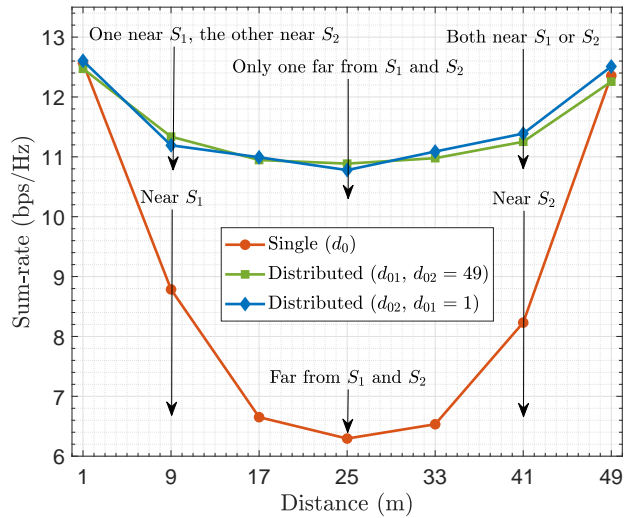}
    \caption{RIS deployment investigation.}
    \label{fig:varyd}
\end{figure}

Figure \ref{fig:varyd} studies the RIS deployment problem in both single and distributed RIS-assisted FD system. In the single RIS {scheme}, the sum-rate gradually increases when the RIS gets closer to $S_1$ or $S_2$. In the distributed RIS {scheme}, two cases are considered: {varying $d_{01}$ when $d_{02} = \unit[49]{m}$ and varying $d_{02}$ when $d_{01} = \unit[1]{m}$.} As both RISs get near the ends or if one is fixed near $S_1$ and the other is near $S_2$, the sum-rate increases. It is shown that when the RIS is located relatively far from both $S_1$ and $S_2$ {in the single RIS scheme}, the distributed RIS scheme significantly improves the sum-rate. This is because deploying distributed RISs enables providing {alternative paths when the other RIS experiences a poor quality link.} For the rest of the paper, it is considered that $d_{01} = \unit[1]{m}$ and $d_{02} = \unit[49]{m}$.







\begin{figure}[h]
    \centering
    \includegraphics[scale= 0.45]{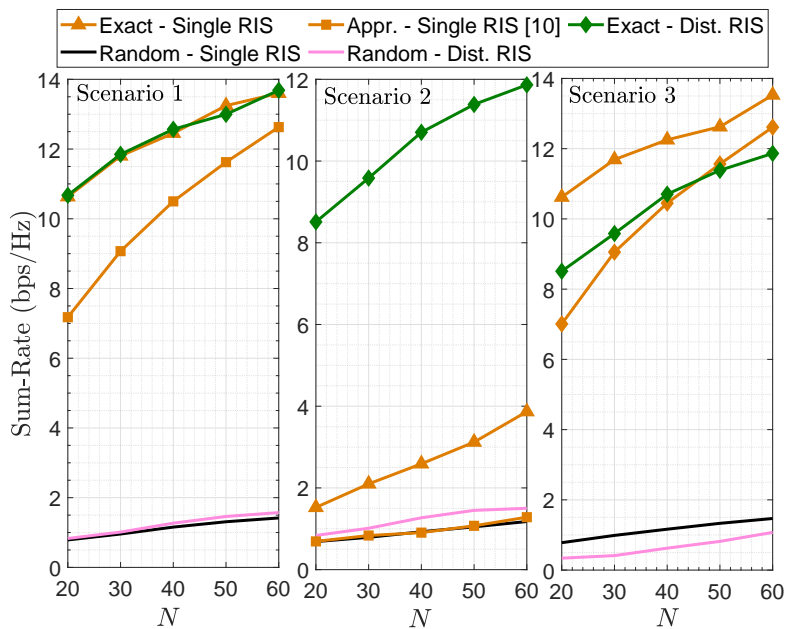}
    \caption{The impact of varying $N$ on the system performance. }
    \label{fig:varyN}
\end{figure}

Figure \ref{fig:varyN} illustrates the effect of increasing $N$ on the system performance. Three practical scenarios are considered to investigate the preference of using single or distributed RIS schemes. In Scenario 1, the distributed and single RIS schemes {achieve} a similar performance due to the strong LoS components (i.e., good quality links), and $N$ is the same in both schemes. In Scenario 2, {the results illustrate} that the distributed RIS system  {significantly} outperforms the single RIS system when the $R_1$-$S_2$ link is blocked/weak. In this case, the distributed RIS scheme has a higher sum-rate since it  {compensates} for the poor quality link by providing an alternative path. On the other hand, if the link between $R_2$-$S_2$ is blocked/week, as in Scenario 3, the single RIS scheme outperforms the distributed RIS since {the former} has double the number of elements compared to the {latter.} It is also worth noting that the proposed DRL algorithm provides a significant improvement in the sum-rate for the single and distributed RIS schemes compared to the random RIS phase shifts in all scenarios. {The performance of the studied scenarios {provides} important insights into the preference of each deployment scheme based on the {link conditions.} Scenario 1 further points that {the} deployment cost should be considered if both schemes yield similar performance,
{as the required channel state information of the single RIS scheme is less than that of the distributed RIS scheme.}}

\begin{figure}[h]
    \centering
    \includegraphics[scale=0.5]{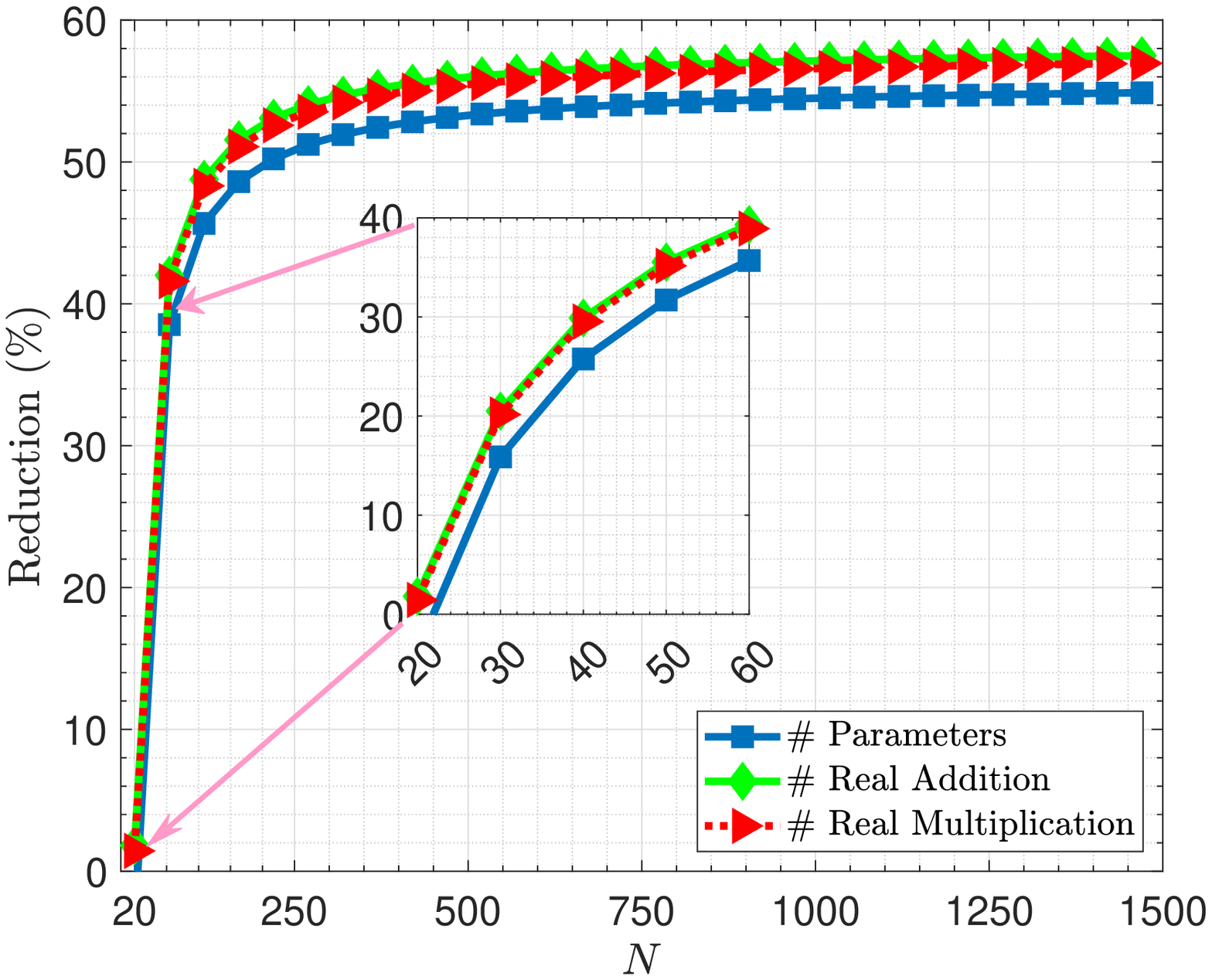}
    \caption{Complexity reduction percentage versus $N$.}
    \label{fig:Comp}
\end{figure}

In the single RIS scheme, the proposed beamforming derivation improves the sum-rate performance in all scenarios, when compared to \cite{A_letter}, as depicted in Fig. \ref{fig:varyN}. Moreover, as shown in Fig. \ref{fig:Comp}, the proposed DRL algorithm provides a complexity reduction percentage up to 40\% for {the} range of $N$ {from 20 to 60} compared to the DRL presented in \cite{A_letter}, and it saturates at 57\% when $N$ is {very} large.

\section{Conclusion} \label{conclusion}
{This letter optimized the  beamformers and RIS phase shifts to maximize the sum-rate for both single and distributed RIS deployment schemes.} Three practical scenarios were considered to investigate the preference of using single or distributed RIS deployment schemes. A closed-form solution is derived to obtain the optimal beamformers, and a novel DRL algorithm is considered for the  {RIS phase shifts optimization. It was shown that the superiority of a deployment scheme depends on the links' quality.} Compared to the non-optimized scenarios, the proposed algorithm significantly  {improved} the sum-rate for both deployment schemes. The proposed DRL algorithm  achieved up to 57\% { complexity reduction} compared to the DRL algorithm in the literature. Future works may consider generalizing the proposed DRL by jointly optimizing the beamformers and RIS phase shifts  for multi-user systems.



\bibliographystyle{IEEEtran}
{\footnotesize\bibliography{IEEEabrv, mybibfile}}
\end{document}